\documentclass[sigconf,anonymous=false]{acmart}

\renewcommand\footnotetextcopyrightpermission[1]{} 
\usepackage{booktabs} 

\setcopyright{rightsretained}

\usepackage{graphicx}

\usepackage{enumitem}
\usepackage{subfigure}

\usepackage{bbm}
\usepackage[colorinlistoftodos]{todonotes}
\usepackage{verbatim} 
\usepackage{tcolorbox} 
\usepackage{graphicx}
\usepackage{tabularx,longtable,multirow,subfigure,caption}
\usepackage{array}
\usepackage{amsmath,amssymb}
\usepackage{url} 

\usepackage{mathtools}

\usepackage{algpseudocode,algorithm,algorithmicx}
\algrenewcommand\algorithmicrequire{\textbf{Precondition:}}
\algrenewcommand\algorithmicensure{\textbf{Postcondition:}}

\begin{document}
\title{Efficient attack countermeasure selection accounting for recovery and action costs}
%


\author{Jukka Soikkeli}
\affiliation{%
  \institution{Imperial College London}
}
\email{jes15@imperial.ac.uk}

\author{Luis Mu\~noz-Gonz\'alez}
\affiliation{%
  \institution{Imperial College London}
}
\email{l.munoz@imperial.ac.uk}

\author{Emil C. Lupu}
\affiliation{%
  \institution{Imperial College London}
}
\email{e.c.lupu@imperial.ac.uk}

\begin{abstract}
The losses arising from a system being hit by cyber attacks can be staggeringly high, but defending against such attacks can also be costly. This work proposes an attack countermeasure selection approach based on cost impact analysis that takes into account the impacts of actions by both the attacker and the defender. 

We consider a networked system providing services whose provision depends on other components in the network. We model the costs and losses to service availability from compromises and defensive actions to the components, and show that while containment of the attack can be an effective defensive strategy, 
it can be more cost-efficient to allow parts of the attack to continue further whilst focusing on recovering services to a functional state. 
Based on this insight, we build a countermeasure selection method 
that chooses the most cost-effective action based on its impact on expected losses and costs over a given time horizon. 
Our method is evaluated using simulations in synthetic graphs representing network dependencies and vulnerabilities, and found to perform well in comparison to alternatives.

\keywords{Countermeasure selection \and Cyber resilience \and Resilience \and Cyber security \and Network security.}
\end{abstract}

\maketitle              

%
%
%
\section{Introduction}
Organisations providing services across the Internet or otherwise connected to an external network face the possibility of cyber-attacks against their systems.
Consequently, such organisations should invest in cyber security to protect their services, both to lower the risk of attacks, and to reduce the impact when they occur.
Determining the desired level for this investment, and its correct application, is not as straightforward as attempting to secure everything fully. The literature on cyber security investment has shown
that a company should never invest to the extent as to cover all potential vulnerabilities or weaknesses \cite{gordon2002economics}, and that a company should retain some part of the security investment budget for when an attack event has taken place \cite{gordon2003information,chronopoulos2018options}. Furthermore, given the existence of unknown vulnerabilities and exploits \cite{njilla2017cyber}, 
not all eventualities can be prepared against. Additionally, security actions can cause limitations to availability, e.g. reductions to communication between systems or users, or loss of compatibility between software applications due to patching. 
Combined, these observations lead to a concern over the cyber-resilience of the system consisting of services the organisation provides and the related network, i.e. the ability of the system to withstand and recover from cyber attacks. 



Given the possibility that an attack against an organisation's system will occur, some cyber-security investment should be allocated into improving the ability of the system to cope during an attack, and recover to normal functionality.
The actions organisations take to mitigate the effects of attacks is the subject of the literature in attack countermeasure selection, reviewed by \cite{Nespoli2017}. 
In contrast to existing work on countermeasure selection, we take a longer term view by focusing on network resilience, forcing the countermeasure selection to consider impacts over time, and recovery dynamics. 

We propose a method for countermeasure selection based on cost impact assessment of both attacker and defender actions, in a medium-to-long-run setting including recovery dynamics. The impacts of actions are estimated using an approach that builds on the attack impact analysis approach by \cite{Albanese2017,Albanese2011}, which uses attack graphs (AGs) and dependency graphs (DGs) to estimate attack impact. Our work adds a cost structure, specifying costs for loss of node availability, the countermeasures, and recovery actions. This extends the impact assessment beyond attack impact alone, including the cost impact of defender's actions alongside that from attack steps for a more general analysis. The structure enables dynamic semi-automatic countermeasure selection based on the overall costs of alternative countermeasure and recovery strategies. This allows our method to make more nuanced countermeasure decisions, balancing attack containment and focus on recovery according to the situation to improve the efficiency of countermeasure strategies.

Our main contribution on the countermeasure selection side, beyond the use of a novel impact assessment approach, is a method to evaluate the effectiveness of each countermeasure based on its expected impact on the system. This includes considering the effects of the countermeasure on the possible evolution of the attack, and on the network's provision of services, both immediately and in a longer term. We achieve this by using the network attack graph to form expectations of possible attack paths and their likelihoods, and using the impact assessment method to estimate the cost impact of the different states the system could end up in which are within a few steps from the current state. 

We test our method by simulations on a small sample graph and on synthetic graphs, against two alternative methods. 
The results suggest that our method provides more cost-efficient countermeasure selection than the alternatives tested, even when the average service performance is close. As our method builds on the attack impact assessment by \cite{Albanese2017,Albanese2011}, we compared our method to an alternative that employs the principles of their attack impact assessment without our extensions. The comparisons show that for the purposes of countermeasure selection, our extensions to their analysis add considerable performance improvements in terms of cost effectiveness and average service performance. 
 
In summary, our contributions include: 1. Extending the attack impact assessment model by \cite{Albanese2017,Albanese2011} into a more general model for impact assessment including defender actions, via modelling service losses and the costs of the countermeasures, and adding a change to enable the modelling of recovery, which the original formulation does not support; 2. Introducing an approach to countermeasure selection based on estimating the expected impacts of actions; 3. Showing that considering recovery and the costs of actions over time can yield a more efficient countermeasure selection.

The structure of the paper is as follows: Sec. \ref{sec:RelatedWork} discusses related work. Sec. \ref{sec:impact_modelling} introduces our impact analysis graph model, including concepts and definitions from \cite{Albanese2017} which our model builds upon.  Sec. \ref{sec:methods} introduces our approach to countermeasure selection in a network. Sec. \ref{sec:evaluation} evaluates the method against two alternative approaches, using simulations. Finally, Sec. \ref{sec:conclusion} concludes. 



\section{Related work} \label{sec:RelatedWork}

In a general sense, \textbf{resilience} is the ability of a system (an organism, a network, a country) to withstand and recover
from adverse events such as natural disasters, epidemics, system faults or cyber attacks. 
Due to the wide use of the term in different fields of study, various definitions for the term have been used \cite{hosseini2016review}.
In the definition we shall use, the defining characteristic of resilience is the \textit{recovery and adaptation} exhibited by a system during and after an adverse event. The systems ability to simply weather an event, which we call \textbf{robustness} after \cite{Ganin2016operational}, covers only part of resilience in our definition. 

A key work in the field of resilience measurement, 
Bruneau et al. \cite{bruneau2003framework} introduced a way to measure resilience based on a system performance curve.
Their method involves calculating resilience loss, the performance lost due to lack of resilience. This is given by difference between the level of a performance metric and its optimal value over time, from the start of the disruption to when performance has recovered fully. 
While alternative measures have also been proposed, such as a method combining two metrics for measuring the resilience of a backbone network by Sterbenz et al.  \cite{sterbenz2010resilience,sterbenz2013evaluation}, 
approaches based on performance curves such as \cite{bruneau2003framework} and its variations have become common in the literature for quantifying system resilience, used in works such as \cite{Ganin2016operational,Yodo2018enabling_control_theory}.
Our work differs from most of the existing literature on resilience by focusing on the actions and investment choices during an ongoing event (attack) and the recovery phase, instead of preparatory planning and capability investment. This focus is intended to address the evolving nature of systems, and adaptation to conditions such as loss of confidentiality or network unavailability etc. 
Most cyber resilience works have focused on the planning and design stage, such as \cite{sterbenz2010resilience,Zhang2015optimization_complex_supply,Ganin2016operational,Shatto2017graph_energy_measure}. Additionally, papers considering reactive response and recovery apply to narrow settings which do not apply to our work. For example, the approaches by \cite{Choudhury2015action_for_cyber,Yodo2018enabling_control_theory} only apply to settings where a control action to correct for a deviation from desired performance is easy to determine in advance, and to apply automatically.

In the cyber security literature, \textbf{countermeasure selection} refers to approaches to choose actions to counter security events (cyber attacks) \cite{Nespoli2017}.
These methods focus on defense against attack events, and as such mainly involve the stages before and during an attack: modelling the system, and the potential attacks and countermeasures; identifying the attack and making a countermeasure choice. We believe that by introducing longer-term resilience considerations into this mix, countermeasure investments can be made with higher efficiency. 

Recent frameworks for attack countermeasure selection 
have commonly used graphical models, including attack graphs \cite{Nespoli2017}. 
The combined use of attack graphs (AGs) and dependency graphs (DGs) for analysing the impact of cyber attacks has been explored by Albanese et al. \cite{Albanese2011,Albanese2017}. Additionally, Kotenko and Doynikova 
\cite{Kotenko2015,Kotenko2016,Doynikova2015} use these graph tools as a foundation for a countermeasure selection system.
Both the approaches connect nodes in the attack graph with relevant parts of the DG information to quantify the impact of a given attack step, but differ in how the DGs are used. 

The method in \cite{Albanese2011,Albanese2017} holds the DG alongside the AG as part of the dynamic analysis, with the effects of attack steps on the DG held as key for impact assessment. By contrast, \cite{Kotenko2014} focuses on enriching the system's AG with topological and service dependency information, and using this augmented AG for the analysis. The DG is employed as a source of component importance information at the pre-processing stage. 
In this work, we shall follow the approach in \cite{Albanese2011,Albanese2017} of using the DG 'live' during the analysis. This is because for our approach to countermeasure choice and resilience analysis, the dynamic effects from attack steps on the dependencies provide useful information beyond that obtained by using the dependency information statically.

Our proposed approach has similarities to the countermeasure selection technique in \cite{Kotenko2015} in the use of AGs and DGs, using costs to quantify attack impact, and basing countermeasure decisions on costs. 
The key differences include our longer-term focus, explicit time dynamics as opposed to comparative statics, and our approach to cost and component valuation via the impact on the final services as opposed to giving each component an intrinsic value.
\section{Impact analysis modelling} \label{sec:impact_modelling}

\subsection{Attack and dependency graphs}
Attack graphs are a network risk-assessment tool that provide a graphical representation of actions that an attacker can take to reach an attack goal, for example root access on a given server, by exploiting vulnerabilities that exist in the network. 
Various types of attack graphs have been proposed in the literature, depending on the application. The work presented here will use the definition of Vulnerability Dependency Graph from \cite{Albanese2017}, a compact AG representation used in their work on attack impact assessment. 
{\small
\begin{definition}{\textbf{(Vulnerability Dependency Graph, \cite{Albanese2017})}}
\textit{
Given a set of vulnerability exploits $V$, a set of security conditions $C$,
a require relation $R_r \subseteq C \times V$, and an imply relation $R_i \subseteq V \times C$, a vulnerability dependency graph G is the directed graph $G = (V,R)$, where $R = \{f(v_i, v_j) \in V \times V : \exists c \in C, (v_i, c) \in R_i \land (c, v_j) \in R_r\}$ is the edge set.
}
\end{definition}
}
\noindent
The vertices consist of nodes representing vulnerability exploits such as "remote exploit of vulnerability V on host A", while the edges implicitly contain security conditions such as "user access to host A".
A require relation $R_r$ between a security condition $c$ and a vulnerability exploit $v_i$ means that $c$ must be satisfied for $v_i$ to be exploited, and an imply relation $R_i$ between $v_j$ and $c$ means that exploit of $v_j$ leads to condition $c$ being satisfied \cite{Wang2006}.
The edges in the set $R$ link pairs of vulnerabilities that are connected via a security condition by what \cite{Albanese2011} call a "prepare-for" relation. A prepare-for relation exists between $v_i$ and $v_j$ if $v_i$ has an edge to security condition $c$ representing an imply relation (exploit of $v_i$ implies the attainment of security condition $c$), and $c$ has an edge to $v_j$ representing a require relation (the exploit of $v_j$ requires the condition $c$ to have been obtained by the attacker). 
This AG representation leaves the security conditions implicit, as the edges are defined as going through a condition but these are not shown, with the end result resembling a dependency graph of the vulnerabilities. 

Dependency graphs (DGs) represent dependency relations between the various components of a system. For example, a server may require input from various databases to perform its function, which themselves may depend on other components.
The DGs in this work represent the \textbf{availability} dependencies of the services provided by software applications in a network. At a given time $t$, the availability of a given node shall be measured as the service level provided at time $t$ as proportion of the service level normally expected of the application, so that full service is represented by 1, while 0 means the service is unavailable. 
The DG nodes represent software applications that provide services across the network, such that applications elsewhere in the network are dependent on them in order to function fully. Dependencies between services are represented by directed edges between the DG nodes. 
For our DGs, we use the definition of a Generalised Dependency Graph from \cite{Albanese2017}: 
{\small
\begin{definition}{\textbf{(Generalised Dependency Graph, \cite{Albanese2011,Albanese2017})}}
\label{def:genDG}
\renewcommand*{\thefootnote}{\fnsymbol{footnote}}
\textit{
A generalised dependency graph is a labeled directed acyclic graph $D=(H,Q,\phi)$, where:
    H is a set of nodes, corresponding to network components;
    $Q=\{(h_1,h_2) \in H \times H : h_1 \text{ depends on } h_2\}$ is a set of edges;
    $\phi : H \rightarrow \mathcal{F}$ is a mapping that associates with each node $h \in H$ a function $f \in \mathcal{F}$ s.t. the arity of $f$ is equal to the outdegree of $h$\footnote{"If $h$ is a terminal node in the dependency graph (i.e. it does not depend on any other node), we assume $\phi(h)$ is the constant (0-ary) function 1"\cite{Albanese2017}.}.
For each node $h \in H$, $h\dot{}$ denotes the set of components that depend on $h$ and $\dot{h}$ denotes the set of components $h$ depends on. 
}
\renewcommand*{\thefootnote}{\arabic{footnote}}
\end{definition}
}
The first two points in the DG definition describe the basic structure of the dependency graph (which nodes are dependent on which others), while the mapping $\phi$ describes the type of dependency that each node $h \in H$ has on its supplier nodes $\dot{h}$. 
We consider the same dependency function types as in (\cite{Albanese2011,Albanese2017}):
\begin{equation}
f_r(a_1,...,a_n) = \begin{cases}
					1 &\text{if $\exists i \in [1,n] \; s.t. \; a_i = 1$}\\
					0 &\text{otherwise}
				   \end{cases}
\end{equation}
\begin{equation}
f_d(a_1,...,a_n) = \frac{1}{n} \sum_{i=1}^n a_i
\end{equation}
\begin{equation}
f_s(a_1,...,a_n) = \begin{cases}
					1 &\text{if $a_i = 1 \; \forall i \in [1,n]$}\\
					0 &\text{otherwise}
				   \end{cases}
\end{equation}
where $a_i$ represents the availability value of a network component which the current component is directly dependent on, and $n$ is the total number of components on which the current component is dependent (i.e. for component $h_i$ we have $n = |\dot{h_i}|$). Here, $f_r$ is a redundancy-type dependency (logical OR), $f_s$ is strict dependency on all supplier nodes (logical AND), and $f_d$ means availability of $h$ is the mean of the availabilities of all of its suppliers (degraded availability). For example, in the sample shown in Fig. \ref{fig:IAG_figure}, for $h_T$ we have $f(s(\dot{h_T},t)) = f_s(a_1) = f_s(s(h_C,t))$.

\subsection{Attack impact analysis}

Our impact assessment approach builds on the \textbf{Impact Assessment Graph} (IAG) proposed in \cite{Albanese2017}, with combined use of an AG and a DG for analysing attack impacts.
We proposes a slight simplification relative to \cite{Albanese2017}. Specifically, in the IAG we use an AG without the compromise time-windows feature in the original, as the handling of time in our model makes this feature unnecessary.
{\small
\begin{definition}{\textbf{(Impact Assessment Graph, modified \cite{Albanese2017})}}
\textit{
Given a vulnerability dependency graph $A=(V,R)$ and a generalised dependency graph $D=(H,Q,\phi)$, an impact assessment graph is a 4-tuple $(A,D,F,\eta)$ where:
	$F \subseteq V \times H$;
    $\eta : F \rightarrow [0,1]$ is a function that associates with each pair $(v,h) \in F$ a real number in the $[0,1]$ interval representing the percentage reduction in the availability of network component $h$ caused by vulnerability exploit $v$.
}
\end{definition}
}
\noindent Effectively, the IAG consists of both the AG and the DG in full, and connections in between them in the form of the function $\eta$ that describes how the availability of the components in the DG are affected by vulnerabilities in the AG. 


The \textbf{network status function} \cite{Albanese2017} defines how the status of a node evolves over time with relation to the availabilities of its dependencies.
The definition states that a network status function $s$ for a DG $D$ assigns each DG node $h$ a value in the range $[0,1]$, capped by the node's dependency function $f$ over the statuses of the node's dependencies. That is, the statuses of the components that $h$ depends on define a maximum availability level for $h$, but the status of $h$ can be below this maximum if directly affected by an attack. 
However, the functional form of the network node status function $s(h,t)$ in \cite{Albanese2017} does not allow modelling recovery in a component's status due to the way the previous period's status $s(h,t-1)$ enters the function definition. A bad availability status one period will continually punish the statuses in future periods, even if a recovery or vulnerability patching has taken place. Therefore, we have redefined the node status function to keep track of all vulnerability exploits that are in effect at time $t$ (either newly exploited, or ones that were exploited and their related security conditions remain compromised), each of which may have a direct impact on $h$. 
Our reformulation also introduces a multiplicative interaction between the direct and indirect availability effects, replacing the minimum function in the original. 
In this way, the direct and indirect effects interact to set the effective availability, in contrast to the dependency availability effect only acting as a cap on the effective availability level. 
These changes also enable 
multiple exploits affecting a node $h \in H$ directly, which the original formulation does not support.
Our proposed formulation for the node status function is: 
\begin{equation}
    s(h,t) = f(s(\dot{h},t)) \prod_{v \in V_{e,t}}(1-\eta(v,h)) \label{eq:status_fn}
\end{equation}
where $V_{e,t}$ is the set of AG nodes that are in an exploited state at time $t$; $\eta(v,h) \in [0,1]$ is the availability effect that the exploit of vulnerability $v$ has on component $h$ (0 for no effect, 1 for fully unavailable); $\dot{h}$ is the set of components that $h$ is dependent on; and  $f(s(\dot{h},t))$ is the availability effect on $h$ from its dependencies. Note that a component recovery is represented as the removal of a vulnerability $v$ from the exploited set $V_{e,t}$, while a patching of a vulnerability will remove the vulnerability $v$ from the full set of vulnerabilities $V$.
The status of network component $h$ is composed of two effects on availability: 1. Compromise effect (direct): $(1-\eta(v,h))$, the effect of compromise (vulnerability exploit) in the AG node $v$ which corresponds to component $h$; 
2. Dependency availability effect (indirect): $f(s(\dot{h},t))$, the effect of unavailabilities in the components that $h$ is dependent on, of type $f_r$ (redundancy), $f_d$ (degradation) or $f_s$ (strict dependence).

We use the example from \cite{Albanese2017} to show how our approach differs. 
The sample represents the network of a small organisation with two final services, an online shopping Web Service ($h_A$) and a Mobile Order Tracking app ($h_C$), their local cache databases ($h_B$ and $h_D$ for online shopping and order tracking, respectively), and a separate subnetwork for the internal logic and a central database powering the services. 

\begin{figure}[!h]
\centering
\includegraphics[width=0.9\columnwidth]{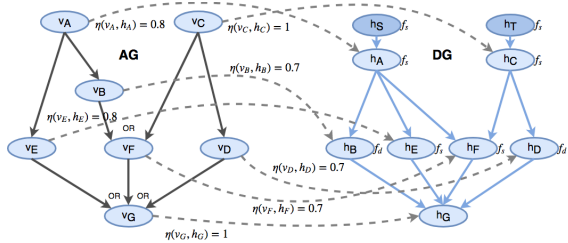}
\caption{\label{fig:IAG_figure}Impact assessment graph, adapted from \protect\cite[Fig. 11]{Albanese2017}}
\end{figure}

The dependency structure and potential attack paths in the sample network are shown in Fig. \ref{fig:IAG_figure}, which recreates the impact assessment graph figure from \cite[Figure 11]{Albanese2017}.
The figure has the AG on the left, and a DG on the right. The nodes in the AG represent vulnerabilities in the different network components. The DG nodes are network components: $h_A$ to $h_G$ provide intermediate services (internal services), while $h_S$ and $h_T$ are customer-facing services (the product of the organisation). The exploit of a vulnerability $v_i$ affects the availability of the corresponding component $h_i$, and the availability of components which are dependent on $h_i$. The dashed lines in Fig. \ref{fig:IAG_figure}
show the availability impact of a vulnerability, $\eta(v_i,h_i)$, for example $\eta(v_F,h_F) = 0.7$ means $h_F$ will lose 70\% of its availability when $v_F$ is exploited (unless the availability of nodes it depends on has decreased, affecting $h_F$ availability). 

The final key component for attack impact analysis using the IAG is the utility derived from the components in the dependency graph: $\forall h \in H, u(h)$ gives the utility for node $h$. The utility is, in effect, the value of the service provided by a given node at each time unit. The work in \cite{Albanese2011,Albanese2017} assume that all the DG nodes are given a utility value which remains static during the analysis. By contrast, we believe that intermediate services provide value only when at least a part of the final service to which they contribute is online (available). Therefore, we only set values for the final service nodes ($h_S$ and $h_T$ in the sample), with the value of the other nodes only reflecting the impact they have on the value arising from the final services. In this way, we use the dependency structure to dynamically determine the values of all intermediate services. 

In contrast to \cite{Albanese2017,Albanese2011}, we propose to introduce a comprehensive
cost model that takes into account the costs of node unavailability, costs of different countermeasures, and recovery costs. Doing so, we aim to move beyond the original \textit{attack} impact assessment to a more general impact assessment by including the estimation of the impacts of defensive actions as well as those of the attacker.  Countermeasure decisions can then made based on expected costs over time, instead of only relying on the projected impacts of attacker actions. 
Our work uses the impact analysis framework as part of an approach countermeasure selection, employing the impact estimates for determining which countermeasure is the most cost effective at a given attack situation. For this purpose, leaving out the analysis of costs of the countermeasure actions could lead to inefficient choices of countermeasures, as the application of a countermeasure can reduce the availability of the services. 

\subsection{Attacks, countermeasures and recovery} \label{sec:attacks}
We model attacks as sequences of attack steps, i.e. atomic exploits of a single vulnerability in an AG, with potential to lead to further steps. They enter the network via particular "entry nodes" which are directly exploitable by the attacker. 
For example, in Fig. \ref{fig:IAG_figure}, an attack can start by an exploit of $v_A$ or $v_C$. 
The potential next steps are then determined by the AG structure -- for example, if $v_A$ was exploited, the possible next steps are $v_B$ and $v_E$. 

The time to compromise a vulnerability $v_i$ (for $i \in V$) is $t_{v_i} = 1$ for all vulnerabilities, so only one attack step can occur during a time unit $t$.
We assume that the probability that an attacker takes an attack step at a given time step is $p_{step}$.
Similarly, we introduce a parameter reflecting the chance that a countermeasure/recovery application is slower than the attacker's next step, $p_{fast-step}$, so the attacker's next step gets executed just before the defender's one. This provides some uncertainty to the timing of events, to avoid limiting us to the case where the defense always beats the attacker to the next step, which seems unrealistic.


\begin{figure*}[!ht]
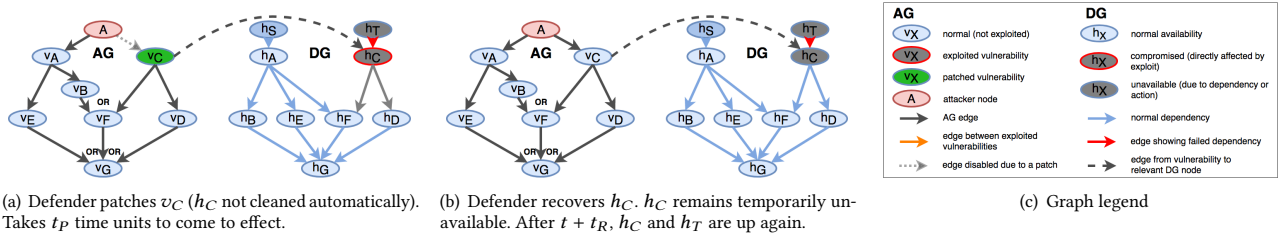
%
\centering
\subfigure[Defender patches $v_C$ ($h_C$ not cleaned automatically). Takes $t_P$ time units to come to effect.]{
  \includegraphics[width=0.3\textwidth]{figures/AG_DG_sample_patch_vCv2.png}
  \label{fig:sample_subfigE}
}\quad
\subfigure[Defender recovers $h_C$. $h_C$ remains temporarily unavailable. After $t+t_R$, $h_C$ and $h_T$ are up again.]{
  \includegraphics[width=0.3\textwidth]{figures/AG_DG_sample_recover_hC_tempv2.png}
  \label{fig:sample_subfigF}
}\quad
 \subfigure[Graph legend]{
  \includegraphics[width=0.3\textwidth]{figures/AG_DG_sample_legendv2.png}
  \label{fig:sample_fig_legend}
}
\caption{Patching and recovery actions and their effects on the system}
\label{fig:sample_patch_recover}
\end{figure*}
\textbf{Countermeasures} (CMs) are actions aimed at reducing the impact of an attack. 
In general these could be of two types, ones affecting the network's security capabilities which can be taken before an attack but not during one (capability changes: redundancy additions, back-ups), and others which can be taken at any time (dynamic countermeasures).
In this paper, we consider one type of dynamic countermeasure, \textbf{patching} vulnerabilities. An another type of dynamic countermeasure, disabling communication links, will be considered in future work, along with capability changes.

The effect of patching is to remove a vulnerability node from the AG, restricting potential attack paths.
A patching action requires $t_P$ units of time to implement, and comes at a cost consisting of a direct cost and a service impact, introduced in Sec. \ref{section:Costs}. 
Fig. \ref{fig:sample_subfigE} illustrates the case where $v_C$ is patched after an exploit. Note that patching does not clear the related DG node $h_C$ from compromise, as recovery is handled separately. 


\textbf{Recovery} refers to actions that the network owner uses to recover the functionality of components (DG nodes) compromised by vulnerability exploits.
To simplify the analysis, we consider the case where there is only one recovery method, which is akin to component replacement.
This consists of a full replacement of a failed or compromised component by a new (clean and working) version with the same functionality (and same vulnerabilities), assumed to take $t_{R}$ time, and cost $c_{R}$. Both the time and cost values are assumed equal across components. 
In terms of the graphs, this corresponds to making a DG node fully functional and the corresponding AG vulnerability not exploited (although it will remain in the AG, so may be compromised later). This is shown in Fig. \ref{fig:sample_subfigF}, showing recovery of $h_C$ after it has been compromised by exploit of $v_C$. 
More detail is provided in Sec. \ref{sec:recovery_process}.

\subsection{Costs of actions} \label{section:Costs}
Our modelling of costs contains direct costs for each action, i.e. node recovery cost $c_{R}$ and patching cost $c_P$, in addition to indirect costs in terms of service loss due to dependencies. 
While the direct costs do matter, the key element of our cost modelling is the loss to the provision of final services caused by component unavailability. \\
\noindent
\textbf{Service loss due to unavailability of DG node $h$ at time $t$:}
\begin{align}
\begin{split}
    g(h,t) = \sum_{h_j \in \textbf{$H_S$}} \Bigl( & u(h_j) \cdot \bigl( f_s(s(\dot{h_j},t) \, | \: s(h,t)=s(h,t-1)) - \\ & f_s(s(\dot{h_j},t) \, | \: s(h,t))\bigr) \Bigr) \label{eq:g_ht}
\end{split}
\end{align}
where we have used $s(h_j,t) = f_s(s(\dot{h_j},t))$ for $h_j \in \textbf{$H_S$}$, which follows from (\ref{eq:status_fn}) and the observation that the customer-facing service nodes $h_S$ and $h_T$ do not have direct exploits (so only the dependency effect counts for them).
The function $g(h,t)$ is the impact of $h$'s deviation, at time $t$, from its previous observed availability level onto the availability of services.

The \textbf{costs of the countermeasure and recovery actions} consist of two parts: the direct cost for the action ($c_{R}$ for recovery of a node, $c_P$ for patching a node, $c_D$ for disabling), and the cost of unavailability of the network components that are directly impacted by the action. For example, the observed (after the fact) cost of patching vulnerability $v_i$ at time $t$ is given by:
\begin{equation}
	c_{patch}(v_i,t) = c_P + \sum_{h_j \in M(v_i)} \left(\sum_{\tau=t}^{t+{t_P}}g(h_j,\tau)\right) \label{eq:observed_c_patch}
\end{equation}
where $v_i$ is a node in the AG, $c_P$ is the direct cost of patching. $M(v_i)$ represents the set of elements in the DG that are adjacent to the AG node $v_i$, that is, the components directly affected by the vulnerability $v_i$. In other words, these are the software where the vulnerability exists, and where the patching of $v_i$ takes place. The current time period is $t$, and $t_P$ represents the time units required for the patching. The inner summation adds together the cost of component $h_j$ being unavailable from $t$ to $t + t_P$. The observed costs due to disabling and node recovery work in a similar manner. 

Note that (\ref{eq:observed_c_patch}) shows the calculation of the observed cost when we know the path of any attack steps and defender actions and therefore know $g(h_j,\tau)$ for $\tau = [t,t+t_P]$. For estimating the cost of an action beforehand, we require an expectation of the state of the model in terms of attacker steps and defender actions during the periods in question. In practice, our approach is to estimate the benefit of an action in terms of an expected trajectory for the system state, 
as explained in Sec. \ref{section:Countermeasure_selection}. 

\subsection{Performance measurement and resilience} \label{sec:resilience_metrics}
We measure the performance of the networked system with an overall service provision status, \textbf{service performance} (SP). It is the weighted sum of the statuses of the client-facing services (the "product" of the organisation), weighted by their relative utilities for the organisation. Mathematically, SP at time $t$ is given as:
\begin{equation}
	SP_t =  \sum_{h \in \textbf{$H_S$}}\frac{u(h)}{\sum_{h_k \in \textbf{$H_S$}} u(h_k)} s(h,t)
\end{equation}
where $s(h,t)$ is the status (availability level) of component $h$ at time $t$, $u(h)$ is the utility the organisation derives from the service component $h$ (during its full availability), and $\textbf{$H_S$}$ is the set of components which are client-facing services. In the sample in Fig. \ref{fig:IAG_figure} 
we have $\textbf{$H_S$} = \{h_S,h_T\}$, as the services are $h_S$ and $h_T$.
When observed over time, this metric can be used to measure the network's resilience based on a performance-curve approach for resilience in the spirit of \cite{bruneau2003framework}. From the curve, resilience metrics can be calculated e.g. according to the approaches by \cite{bruneau2003framework} or \cite{Ganin2016operational}.

\subsection{Recovery process} \label{sec:recovery_process}
We evaluate our model in a setting with automatic recovery decisions, where the choice of whether to recover a node is based on the likely benefit versus the costs of the recovery strategy. 
We keep this recovery process separate from countermeasure choices, which simplifies expectation formation for CM selection. 
Node recovery is done if it leads to a reduction in losses exceeding the recovery cost.
The loss reduction from recovering node $h$ at time $t$ is:
\begin{equation}
    LR(h,t) = Loss_{\neg R}(h,t) - Loss_R(h,t)
\end{equation}
where $Loss_{R}(h,t)$ is the loss with recovery, which is:
\begin{align}
\begin{split}
    Loss_{R}(h,t) = & \sum_{\tau=t}^{t_{max}} \mathbb{E}_{t} (g(h,\tau|s(h,\tau)=0)) + \\
    & \sum_{\tau=t_{max}}^{t_{horizon}}\mathbb{E}_{t}(g(h,\tau|s(h,\tau)=1))
\end{split}
\end{align}
where $t_{max} = t+t_{R}+t_{P}+t_{D}$ is the time it would take to recover, patch and re-enable edges (if necessary), and $t_{horizon}$ is the last time period in the time horizon considered. This metric consists of the loss from service unavailability from time $t$ until $t_{max}$ when the recovery is finished, and from time $t_{max}+1$ onward when the node will be assumed recovered (and not re-compromisable). Loss without recovery, $Loss_{\neg R}(h,t)$, is given by:
\begin{align}
    Loss_{\neg R}(h,t) = & \sum_{\tau=t}^{t_{horizon}}\mathbb{E}_{t}(g(h,\tau|s(h,\tau)=0))
\end{align}

\subsection{Sample impact analysis for CM selection} \label{sec:sample_analysis}
Choosing countermeasures by focusing solely on stopping the progress of an attack
means ignoring: 1. the cost of countermeasure actions (direct and indirect), and 2. recovery of the network toward a desired state. 
Disregard to these aspects can lead to choices that are not efficient in the longer term. 
Assume that the network in Fig. \ref{fig:IAG_figure} experiences an attack exploiting $v_C$ at $t=0$. 
As $v_C$ is exploited, $h_C$ becomes unavailable, and so does $h_T$ due to its dependency on $h_C$.
While the exploit of $v_C$ already causes considerable damage in terms of service losses, once an attacker has exploited $v_C$, it can further exploit $v_F$ or $v_D$. 
While the exploit of $v_D$ only affects $h_D$ (as $h_C$ is down) resulting in no change in either cost or SP, an attack on $h_F$ takes down the remaining service $h_S$, causing full loss of service and cost due to service losses of $u(h_T) + u(h_S)$ per time period. 
When applied repeatedly at each step, the marginal impact analysis approach from \cite{Albanese2017,Albanese2011} would choose to patch the most high-impact component that could be affected by an attack next. 
Given the compromise of vulnerability $v_C$, at the next time step $t=1$, the next attack steps could be $v_F$ or $v_D$, affecting components $h_F$ and $h_D$, respectively. Choosing based on only on the potential impacts of these attack steps, the choice would be to patch $v_F$.
%
Given this choice, at time $t=1$ the attacker could proceed to exploit $v_D$, which was not patched. The appropriate reaction to this would be to patch $v_G$, given another round of marginal impact calculation. 
Finally, to return to full availability of services, nodes $v_D$ and $v_C$ need to be recovered ($v_c$ with patching, $v_D$ without). 

While the above approach is sensible if we want to guarantee that the higest impact nodes $h_F$ and $h_G$ are never compromised, it may not be cost efficient over time. 
The costs of the actions, or recovery, are not considered by the above countermeasure strategy.
However, the choice of actions and their order can have a large impact on the costs, especially those arising from service losses. Note that, as the service impact cost $g(h,t)$ for node $h$ is nonzero only when the node causes a change in the status of the services, the service impact of a given node $h$ can change over time. For example, the unvailability of $h_G$ only impacts services when $h_F$ is available. This changing loss impact can make a great difference on the overall cost of a countermeasure strategy.
For example, if there is a low probability that the attack will have successfully moved to a different node before the next time step, it may be better to \textbf{not} take a countermeasure that contains the attack, but focus on recovering and patching $v_C$. In the worst case, the attacker has been able to move fast enough to exploit $v_F$ before $h_C$ goes offline for patching and recovery. In this case, all services go down at $t=1$, so recovering $h_F$ will have to be done -- but this happens without additional availability impact, as services will be down already. 
Again, in the worst case the attacker may move fast enough to compromise $h_G$ before the system $h_F$ is taken down for recovery, so $h_G$ will require recovery at $t=3$. 
Even in this worst case the overall costs could be lower than with the approach from earlier, if the time to recover is lower than to patch $t_R < t_P$. 
However, this approach also benefits from there being a chance that the attacker will not successfully make another step before the defender reacts, meaning in the best case only $h_C$ has to be recovered and patched. 

This sample highlights that, depending on the situation, it can be more cost effective to act reactively to rely on recovery capabilities, while sometimes a proactively containing the attack can be better. We built our CM selection approach on cost impact analysis to be able to find the approach that works best in a given situation. 

\section{Countermeasure selection} \label{sec:methods} \label{section:Countermeasure_selection}
This section describes our approach to countermeasure selection based on the cost impacts of defender actions in addition to expected attack steps. We shall refer to this method as Cost-Impact Countermeasure Selection, or \textbf{CICM}.
Our CM selection algorithm essentially estimates the impact of all of the countermeasures that apply to an AG node of interest $v_s$, and returns a list of them in descending order of their overall benefit relative to the trajectory that would be expected in absence of a countermeasure. The highest benefit CM is implemented, provided that the benefit is positive.

The effectiveness of each countermeasure is evaluated by comparing the expected costs of the CM to the benefits it is expected to yield. Our approach uses the direct cost of the action, and two potential evolutions ("trajectories") of the system: the "expected trajectory" reflects how an attack would be expected to proceed within the network in the absence of the countermeasure, and measures what the impact would be in terms of costs due to both availability loss and recovery actions. The "deviating trajectory" given a countermeasure action measures the expected costs due to availability loss and recovery when the countermeasure is applied. As a countermeasure initially requires a network component to be taken offline temporarily, we make a difference between the immediate and the longer-run impacts of the countermeasure. Therefore, our measure of the benefit arising from a countermeasure is given by:
\begin{align}
\begin{split}
    B(cm,v_i,t) = & eaf(v_i) \cdot (t_{horizon}-t) \cdot trajD_{LR}(cm,v_i) \\
                                     & + trajD_{curr}(cm,v_i) - c_{cm} \label{eq:CM_benefit}
\end{split}
\end{align}
\noindent
where the first part is an estimate of benefit in later time steps, consisting of the long-run trajectory difference $trajD_{LR}(cm,v_i)$ multiplied by the expected frequency of future attacks exploiting $v_i$, $eaf(v_i)$, and the time periods left until the end of the horizon. The second term is the trajectory difference currently (until the CM has been successfully applied), and the last term is the direct cost of the CM, $c_{cm}$.
The expected attack frequency to node $v_i$, $eaf(v_i)$, 
is an estimate of the probability that the attacker will attempt to exploit node $v_i$ again. We estimate this by approximating the probability of the shortest viable (not patched) path from the attacker node $A$ to $v_i$. This we calculate as the step probability $p_{step}$ to the power of the number of edges on the shortest viable attack path from $A$ to $v_i$. The approximation is by no means fully accurate, but captures the behaviour we want, and is considerably simpler and faster than a rigorous calculation of the exact probability. The exact calculation would be excessive given the amount of uncertainty arising from other parts of the model and its environment. 
Note that this is the probability of an attacker to (re-)obtain the privileges required to exploit $v_i$ in the future, via any path, and not based on the current compromise state.

The current and long-run trajectory differences are given by: 
\begin{align}
    trajD_{curr}(cm,v_i) &= devTraj_{curr}(cm,v_i) - expTraj(v_i) \\
    trajD_{LR}(cm,v_i) &= devTraj_{LR}(cm,v_i) - expTraj(v_i)
\end{align}

\begin{figure}[t]%
\centering
\includegraphics[width=\columnwidth]{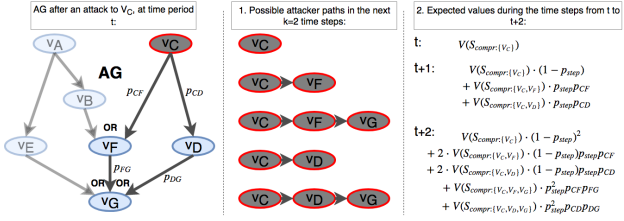}
\caption{The stages of the expected trajectory calculation}
\label{fig:expTraj_sample}
\end{figure}

The expected trajectories $expTraj(v_i)$, $devTraj(cm,v_i)_{curr}$ and $devTraj(cm,v_i)_{lr}$ are calculated by formulating an expected value for availability impact and costs in the next $k$ time steps. For $expTraj(v_i)$, this proceeds as follows: 1. From the AG, estimate what paths the attacker could follow in the next $k$ time steps starting from node $v_i$, the current head of the attack. Estimate the impact of each of these potential paths on service availability and on recovery costs. 2. Formulate expected values for the services impact and costs in each of the time steps from $t$ to $t+k$, using the paths calculated in step 1 with the probabilities for each exploit, the probability for attack step being completed in each time step. An illustration of this is provided in Fig. \ref{fig:expTraj_sample}, for the case where the vulnerability $v_C$ in the sample graph has been exploited at time period $t$. The first panel shows the AG situation at time $t$, the middle panel shows the possible attack paths in the next $k=2$ time steps. The last panel shows the calculation of the expected values for the time periods $t, t+1, t+2$, where $V(S)$ refers to the valuation (service performance or cost) at system state $S$. For brevity, the figure uses a shorthand for states with only the exploited nodes listed, e.g. $S_{compr:\{v_C,v_D\}}$ stands for the state where the set of exploited nodes is $\{v_C,v_D\}$. 
The "deviating trajectory" estimates given a countermeasure $cm$ are calculated similarly to $expTraj(v_i)$, but assuming that a CM is applied. An additional difference is that when calculating the expected values (step 2 above), we further consider the possibility that the CM is not applied in time before the next step by the attacker, represented by the probability parameter $p_{fast-step}$. We calculate two deviating trajectories as the application of a CM causes temporary unavailability at first, before the longer-run effect is obtained.

The split into current and longer term trajectory impacts enables a simultaneous consideration of the immediate up-front costs of a countermeasure (direct cost, temporary availability impact) and the effect it has on the robustness of the network in future periods. 
Our method accounts for the future by calculating the "expected trajectory" impact and the expected number of times such an attack would be observed in the future. This approach provides an estimate of longer term impact that is considerably more efficient to calculate than directly considering all possible states in each future period. 
A weakness of the method is that it considers future impacts relative to the current situation, so excludes the impact of changes to the network or the environment that might happen in the future. However, such information is unlikely to be available at the time when the decision is to be made. In the case it was, it should be possible to establish the robustness of the estimate to such changes, for example by calculating the impact of the CM conditional on other CMs being applied as well. However, for the sake of tractability and scalability, we followed a greedy strategy looking for the impact of the countermeasures independently. 

While we calculate probabilities of various paths of attacks to formulate the expected trajectory, we do not solve for the whole network or the full time horizon, as such an approach would not scale to realistic network sizes. Instead, the method focuses on "neighborhoods" of the current attack, by looking at potential states a few time-steps forward from the current "boundary" of the attack.
As another way to simplify the problem, we don't consider patching at each of the nodes, but focus on the nodes nearest to the entry (if patchable) [for future attacks], and those on the attack "boundary" [for containing the ongoing attack, as well as future attack steps].

\section{Evaluation} \label{sec:evaluation}

We used simulations to investigate the usefulness of the proposed framework for countermeasure selection, by comparison to two alternative strategies. The first comparison point is what we call the Attack Impact Approach (\textbf{AIA}),
which we built by adapting the attack impact assessment approach from \cite{Albanese2017} to be used for automated countermeasure selection. While their original work was not proposed for automated countermeasure selection, 
we built a CM selection approach for patching actions using the main principles of their attack impact assessment method, using their marginal impact metric for patching choices. 
To use this approach in a setting where we care about performance over time, we also required the method to handle recovery, so included our version of the node status calculation in AIA. 
AIA can be considered a containment approach, as it chooses patching actions that apply to the vulnerabilities exploitable next by the attacker, not ones that apply to the already exploited vulnerabilities. This has the potential for stopping the attacker from compromising important nodes, but applying CMs on healthy components will lead to temporary availability losses, which can be costly.

The second comparison is to a strategy where a patch is always applied to the latest exploited vulnerability, without considering costs or alternative actions. We call this \textbf{PLE}, for "Patch Latest Exploit". This approach rejects containment in favour of blocking the last used attack paths from being exploitable in the future, effectively limiting future exposure while accepting current risk. PLE benefits from limiting the availability loss from CM actions, as actions are only applied to nodes that already suffer from reduced availability. While the two comparison approaches represent extremes in terms of containment and treatment focus, our approach is intended to be able to choose a cost-efficient approach somewhere in between these extremes based on the attack situation. In the simulations, the same recovery recovery process, described in section \ref{section:Countermeasure_selection} is used for all of the approaches.

We ran tests under two settings: the sample graph in Fig. \ref{fig:IAG_figure}, and randomly generated synthetic graphs. The sample graph provides a useful basis for comparisons as the attack impact assessment method by \cite{Albanese2017,Albanese2011} was demonstrated on it. The generated graphs are directed acyclic graphs of a specified size (in terms of nodes). For simplicity, the number of AG nodes was restricted to match that of the DG nodes. To control the structure, we have restricted the maximum number of parents of a node (nodes dependent on the node) to three. 
The connections between the AG and DG nodes are chosen at random, meaning that the vulnerabilities in the AG correspond to random system components (in DG), and the attack paths on the AG can be considerably different from paths in the dependency structure. 
Furthermore, in the DG, the number of dependencies (children) of each node are chosen at random, as are the dependency functions for each node. For each DG, two nodes are allocated as service nodes. In the AG, the number of children of a given node (number of further vulnerability exploits made possible by a given exploit) is drawn at random. To introduce attack entry points, we add a node representing the attacker's starting point, and its children drawn among the other nodes are the entry nodes. Additionally, probability values for the AG edges are set, representing the ease of exploiting a node, drawn from a distribution corresponding to the access complexity metric of CVSS scores \cite{firstCVSS}.

We simulated randomly generated attacks into each graph in question. 
The attacks follow a path towards a goal node, which is picked from among the AG nodes with the highest availability impact on the final services (the single highest impact one, or drawn among the shared highest impact nodes). Each chosen edge is picked from those along the paths to the goal, based on a draw between viable candidates, where the distribution is based on the edge probability values (representing access complexity). This creates variety across simulations, approximating different attacker choices based on e.g. different skill levels. One vulnerability exploit is allowed per time period. If a given step is not possible, due to a CM action, the attacker attempts another exploit that is on a path to the goal. If the goal becomes unreachable, the attack will stop. 



Unless otherwise stated, the simulations use the following parameter values: one unit of time to compromise a vulnerability $t_{v_i} = 1$, two units to patch a vulnerability ($t_P = 2$), and one to recover a node $t_R = 1$. 
The direct costs are $c_P = 2$ for patching, and $c_R = 3$ for node recovery. 
There is a 30\% probability that an attacker takes an attack step at a given time step ($p_{step} = 0.3$), and $p_{fast-step} = 0.3$, so there is a 30\% chance that a given CM/recovery application is slower than the attacker's next step.

\subsection{Results for the sample graph}

\begin{figure}[t]%
\centering
\subfigure[Resilience curve (SP); fully detected steps]{
  \includegraphics[width=0.45\columnwidth]{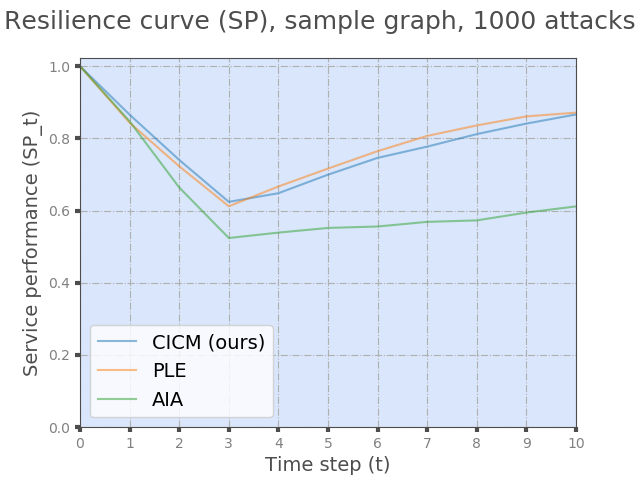}
}\quad
\subfigure[Costs and service losses over time; fully detected steps]{
  \includegraphics[width=0.45\columnwidth]{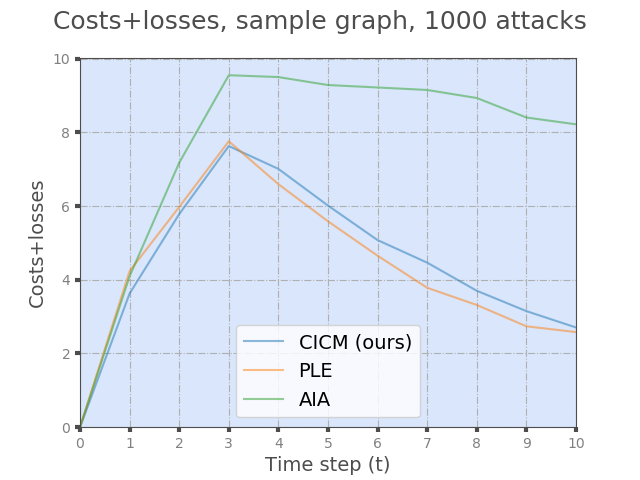}
}\quad
\subfigure[Resilience curve (SP); 1 undetected step]{
  \includegraphics[width=0.45\columnwidth]{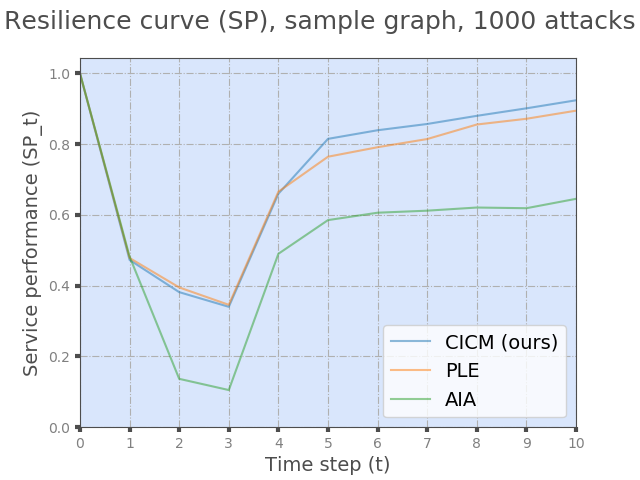}
}\quad
\subfigure[Costs and service losses over time; 1 undetected step]{
  \includegraphics[width=0.45\columnwidth]{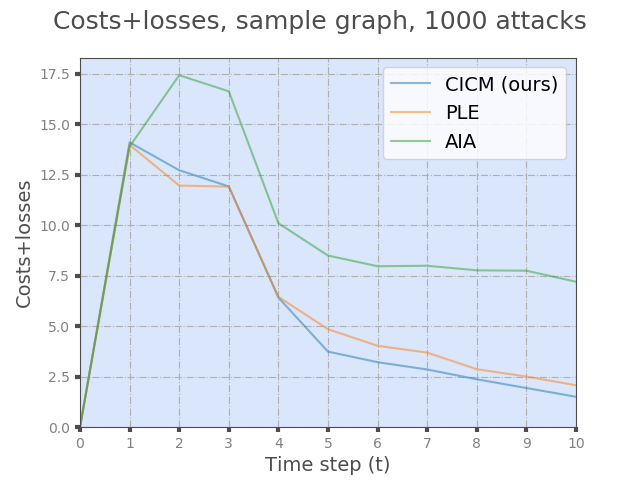}
}
\caption{Simulation results for the sample graph}
\label{fig:Sample_graph_simul_Simple_AJ}
\end{figure}

The simulation results on the sample graph are shown in Fig. \ref{fig:Sample_graph_simul_Simple_AJ}. The figure shows a comparison of our method to the two alternative CM schemes, both for the resilience curve (SP metric), and for costs and service losses over time. The curves represent the mean values for the metrics (SP, overall costs) across 1000 simulated attacks; the blue curve ("CICM") indicates our method, the orange one is for PLE and the green for AIA. 
The two upper panels show simulations where the attacks are detected immediately at the first step, and the defender actions (countermeasures, recovery) can be started immediately. By comparison, on the lower two panels, there is a delayed detection of the initial attack step, in which the attacker has already done one step before a step is detected (that is, the second step overall) and the defender actions start. We believe that this case is more close to reality, as attacks can remain undetected in a network for long periods of time. 

Comparing the results for our approach to those from the AIA alternative, we notice our method outperforming AIA both with regard to service performance over time, and the overall costs. 
The difference between our method and AIA becomes large early on and fails to recover afterwards. 
This is true whether we look at the case with immediate attack detection (upper panels), or the case with delayed detection (lower panels). Using our approach, the mean SP was over 20\% better (23\% for immediate detection, 27\% for delayed), and the overall costs were 42\% lower on average than for AIA (for both immediate and delayed detection). 
The Wilcoxon signed-rank test rejected the hypothesis of the difference being zero for SP and overall costs in both immediate and delayed case, suggesting that the performance difference is statistically significant.
The result appears to be due to a reduced service availability, which causes the similarity in the SP and overall costs (of which service losses is a part). As these are average curves over randomised attacks, some of the detail related to the individual runs is averaged out, but it seems that AIA has a harder time purging the attack from the system than the other approaches, so the difference to other approaches remains until the end of the time window shown.

The PLE approach exhibits performance roughly matching ours, both when the detection of the attack is instant (upper panels), and when there is a delay (lower panels). 
In fact, the difference between the approaches was found statistically insignificant using the Wilcoxon test. 
In the case of this graph, this result can be expected. 
As discussed in Sec. \ref{sec:sample_analysis}, to obtain good CM selection performance, the initial exploit will always be patched in this sample graph as the services are directly and fully dependent on the entry nodes. 
Additionally, the small size of the sample graph limits the variety in attacker steps, and thus on the CM choices of the defender.

\subsection{Results for randomly generated graphs} \label{sec:sim_gen_graphs}
Simulations were ran on graphs of different sizes, and varying the detection delay between no delay and a delay of two steps. The size classes vary the number of nodes in the DG (10, 20 and 50-node DGs), with a corresponding number of AG nodes. We generated 100 graphs of each size, and ran 100 random attacks on each graph. 

Table \ref{table:evaluation_table_gen_AJ} shows the results for comparisons to the AIA approach in the case of immediate detection. The results for delayed detection are overall very similar, so we have omitted them to save space.  We found a considerable benefit for CICM in terms of both the SP metric and overall cost. While the size of the difference in overall costs between the approaches gets smaller with the size of the graphs, due to a general improvement of both of the approaches, the relative cost saving for CICM is around 53-54\% across graph sizes for immediate, and 46-51\% for delayed detection.
Using the Wilcoxon signed-rank test, 
we find that CICM demonstrated statistically significantly better results than AIA for both of the metrics.

\begin{table}[t]
\noindent
\centering
\caption{Comparing mean values of performance metrics, our approach (CICM) vs AIA; immediate detection}
\resizebox{0.47\textwidth}{!}{
 \begin{tabular}{ >{\centering\arraybackslash}p{1.0cm}|| 
 >{\centering\arraybackslash}p{0.6cm}|
 >{\centering\arraybackslash}p{0.6cm}|
 >{\centering\arraybackslash}p{0.6cm}||
 >{\centering\arraybackslash}p{0.6cm}|
 >{\centering\arraybackslash}p{0.6cm}|
 >{\centering\arraybackslash}p{0.6cm}|
 >{\centering\arraybackslash}p{0.6cm}|
 >{\centering\arraybackslash}p{0.6cm}|
 >{\centering\arraybackslash}p{0.6cm}  } 

 \multicolumn{1}{c}{\textbf{}} & \multicolumn{3}{c}{\textbf{CICM}} & \multicolumn{6}{c}{\textbf{Difference: CICM - AIA}} \\
 \hline
 \multirow{2}{*}{\textbf{\small{DG size}}}	& \multirow{2}{*}{\small{10}} & \multirow{2}{*}{\small{20}} & \multirow{2}{*}{\small{50}} & \multicolumn{2}{c|}{\small{10}} & \multicolumn{2}{c|}{\small{20}} & \multicolumn{2}{c}{\small{50}} \\
 \cline{5-10}
 \textbf{}	&  &  &  & \scriptsize{diff.} & \scriptsize{\# +/-} & \scriptsize{diff.} & \scriptsize{\# +/-} & \scriptsize{diff.} & \scriptsize{\# +/-} \\
 \hline

  SP       & \small{0.862} \scriptsize{(0.301)} & \small{0.904} \scriptsize{(0.243)} & \small{0.946} \scriptsize{(0.169)}
  			& \small{0.168} \scriptsize{(0.352)} & \scriptsize{95/3} 
            & \small{0.132} \scriptsize{(0.298)}& \scriptsize{95/2} 
            & \small{0.077} \scriptsize{(0.224)}& \scriptsize{100/0}\\ 
\scriptsize{p-value} & \scriptsize{-} & \scriptsize{-} & \scriptsize{-} 
			& \scriptsize{\textbf{0.00}} & \scriptsize{-} 
            & \scriptsize{\textbf{0.00}} & \scriptsize{-} 
            & \scriptsize{\textbf{0.00}} & \scriptsize{-} \\ 

    Cost    & \small{3.49} \scriptsize{(7.14)} & \small{2.71} \scriptsize{(5.92)} & \small{1.84} \scriptsize{(4.32)}
    		& \small{-3.91} \scriptsize{(8.15)} & \scriptsize{4/96} 
            & \small{-3.25} \scriptsize{(7.00)}& \scriptsize{0/100}  
            & \small{-2.17} \scriptsize{(5.33)}& \scriptsize{0/100} \\ 
\scriptsize{p-value} & \scriptsize{-} & \scriptsize{-} & \scriptsize{-} 
			& \scriptsize{\textbf{0.00}} & \scriptsize{-} 
            & \scriptsize{\textbf{0.00}} & \scriptsize{-} 
            & \scriptsize{\textbf{0.00}} & \scriptsize{-} \\ 
 \hline
 \hline
 \multicolumn{10}{l}{\scriptsize\raggedright{Notes: 100 graphs per size, 100 attack simulations each; "\# +/-": count of positive/negative diffs.}}
\end{tabular}
}
\label{table:evaluation_table_gen_AJ}
\end{table}

\begin{table}[t]
\noindent
\centering
\caption{Comparing mean values of performance metrics, our approach (CICM) vs PLE}
\resizebox{0.47\textwidth}{!}{
 \begin{tabular}{ >{\centering\arraybackslash}p{1.0cm}|| 
 >{\centering\arraybackslash}p{0.6cm}|
 >{\centering\arraybackslash}p{0.6cm}|
 >{\centering\arraybackslash}p{0.6cm}||
 >{\centering\arraybackslash}p{0.6cm}|
 >{\centering\arraybackslash}p{0.6cm}|
 >{\centering\arraybackslash}p{0.6cm}|
 >{\centering\arraybackslash}p{0.6cm}|
 >{\centering\arraybackslash}p{0.6cm}|
 >{\centering\arraybackslash}p{0.6cm}  } 

 \multicolumn{10}{c}{\textbf{Immediate attack detection}} \\
 \hline
 \multicolumn{1}{c}{\textbf{}} & \multicolumn{3}{c}{\textbf{CICM}} & \multicolumn{6}{c}{\textbf{Difference: CICM - PLE}} \\
 \hline
 \multirow{2}{*}{\textbf{\small{DG size}}}	& \multirow{2}{*}{\small{10}} & \multirow{2}{*}{\small{20}} & \multirow{2}{*}{\small{50}} & \multicolumn{2}{c|}{\small{10}} & \multicolumn{2}{c|}{\small{20}} & \multicolumn{2}{c}{\small{50}} \\
 \cline{5-10}
 \textbf{}	&  &  &  & \scriptsize{diff.} & \scriptsize{\# +/-} & \scriptsize{diff.} & \scriptsize{\# +/-} & \scriptsize{diff.} & \scriptsize{\# +/-} \\
 \hline

  SP       & \small{0.862} \scriptsize{(0.301)} & \small{0.904} \scriptsize{(0.243)} & \small{0.946} \scriptsize{(0.169)}
  			& \small{0.001} \scriptsize{(0.112)} & \scriptsize{42/19} 
  			& \small{0.000} \scriptsize{(0.110)} & \scriptsize{38/30} 
            & \small{0.000} \scriptsize{(0.080)}& \scriptsize{29/38}\\ 
\scriptsize{p-value} & \scriptsize{-} & \scriptsize{-} & \scriptsize{-}
            & \scriptsize{\textbf{0.01}} & \scriptsize{-} 
            & \scriptsize{0.17} & \scriptsize{-} 
            & \scriptsize{\textbf{0.00}} & \scriptsize{-} \\ 

    Cost    & \small{3.49} \scriptsize{(7.14)} & \small{2.71} \scriptsize{(5.92)} & \small{1.835} \scriptsize{(4.32)}
    		& \small{-0.03} \scriptsize{(2.69)} & \scriptsize{34/58} 
            & \small{-0.03} \scriptsize{(2.59)} & \scriptsize{43/56} 
            & \small{0.04} \scriptsize{(2.23)}& \scriptsize{36/64} \\ 
\scriptsize{p-value} & \scriptsize{-} & \scriptsize{-} & \scriptsize{-} 
			& \scriptsize{\textbf{0.00}} & \scriptsize{-} 
            & \scriptsize{\textbf{0.00}} & \scriptsize{-} 
            & \scriptsize{\textbf{0.02}} & \scriptsize{-} \\ 
 \hline
 \hline
 
 \multicolumn{10}{c}{\textbf{Delayed attack detection, two undetected steps}} \\
 \hline
 \multicolumn{1}{c}{\textbf{}} & \multicolumn{3}{c}{\textbf{CICM}} & \multicolumn{6}{c}{\textbf{Difference: CICM - PLE}} \\
 \hline
 \multirow{2}{*}{\textbf{\small{DG size}}}	& \multirow{2}{*}{\small{10}} & \multirow{2}{*}{\small{20}} & \multirow{2}{*}{\small{50}} & \multicolumn{2}{c|}{\small{10}} & \multicolumn{2}{c|}{\small{20}} & \multicolumn{2}{c}{\small{50}} \\
 \cline{5-10}
 \textbf{}	&  &  &  & \scriptsize{diff.} & \scriptsize{\# +/-} & \scriptsize{diff.} & \scriptsize{\# +/-} & \scriptsize{diff.} & \scriptsize{\# +/-} \\
 \hline

  SP       & \small{0.801} \scriptsize{(0.361)} & \small{0.852} \scriptsize{(0.302)} & \small{0.901} \scriptsize{(0.235)}
  			& \small{0.044} \scriptsize{(0.235)} & \scriptsize{78/19} 
            & \small{0.034} \scriptsize{(0.197)} & \scriptsize{71/21} 
            & \small{0.020} \scriptsize{(0.154)}& \scriptsize{68/25}\\ 
\scriptsize{p-value} & \scriptsize{-} & \scriptsize{-} & \scriptsize{-} 
			& \scriptsize{\textbf{0.00}} & \scriptsize{-} 
            & \scriptsize{\textbf{0.00}} & \scriptsize{-} 
            & \scriptsize{\textbf{0.00}} & \scriptsize{-} \\ 
            
    Cost    & \small{5.04} \scriptsize{(8.86)} & \small{4.16} \scriptsize{(7.60)} & \small{3.33} \scriptsize{(6.32)}
    		& \small{-1.09} \scriptsize{(5.60)} & \scriptsize{11/86} 
            & \small{-0.96} \scriptsize{(4.85)} & \scriptsize{8/91} 
            & \small{-0.80} \scriptsize{(4.08)}& \scriptsize{6/94} \\ 
\scriptsize{p-value} & \scriptsize{-} & \scriptsize{-} & \scriptsize{-} 
			& \scriptsize{\textbf{0.00}} & \scriptsize{-} 
            & \scriptsize{\textbf{0.00}} & \scriptsize{-} 
            & \scriptsize{\textbf{0.00}} & \scriptsize{-} \\ 
 \hline
 \hline
 \multicolumn{10}{l}{\scriptsize\raggedright{Notes: 100 graphs per size, 100 attack simulations each; "\# +/-": count of positive/negative differences.}}
\end{tabular}
}
\label{table:evaluation_table_gen_Simple_combined}
\end{table}

Comparisons to the PLE strategy are displayed in Table \ref{table:evaluation_table_gen_Simple_combined}.
When there is no delay in attack detection, our approach performs, on average, almost identically to PLE strategy in terms of the performance metric SP. The difference values are practically zero, and the Wilcoxon test fails to reject the null hypothesis of equal means for SP for the 20-node case. On the side of overall cost, the difference magnitudes are also small. While the simulations show CICM slightly outperforming PLE on average for 10 and 20-node graphs, with 50-node graphs the results are mixed. Overall, CICM was more cost-efficient in 64\% of the 50-node graphs simulated, but the mean difference over all graphs is in favour of PLE (2\% cost difference). The main message here is: if the attack is detected immediately at the point of entry, there is little difference between CICM and PLE. 
This makes sense, as patching the first node stops the attacker from regaining access to the network after the attack is purged elsewhere, and immediate detection makes this very effective.

When there is a delay in initial detection of the attack, the benefits of CICM over PLE become clear. 
There is some evidence of a benefit on the average service performance relative to PLE (amounting to between 0.2\% and 0.5\% of the overall value of the output of the services, on average). 
Regardless, the main impact is on the overall cost side, where our method provides average efficiency improvements over PLE amounting to 18\% for 10-node, 19\% for 20-node and 20\% for 50-node graphs. Importantly, the cost savings are consistent, with benefits obtained in 90\% of all the graphs tested (see the +/- counts in the lower part of Table \ref{table:evaluation_table_gen_Simple_combined}: 86\% of 10-node graphs had a negative sign for the cost difference, 91\% of 20-node graphs, and 94\% of 50-node).   
The fact that the difference shows up in cost instead of SP makes sense, as our method chooses actions based on the overall cost, not on SP. The results also show that this effect grows with the size of the graph, suggesting that larger graphs provide more room for choices that improve cost efficiency. 

Summarising the findings from Table \ref{table:evaluation_table_gen_Simple_combined},
we conclude that while a straightforward patching strategy like PLE can yield good results in terms of service performance, our method provides considerable cost savings when attacks are detected with a delay. 

We investigated the sensitivity of the results to the length of delay in detection. We varied the number of attack steps that go undetected before the defender starts their CM selection and recovery processes, with the other parameter values held constant at the levels mentioned above. For this, we used the 20-node generated graphs (100 graphs, 100 simulated attacks for each graph), comparing to PLE. 
The results show that the cost savings from CICM relative to PLE increase drastically when moving from immediate to delayed detection, with the saving jumping from 1.1\% for immediate detection (0 undetected steps) to 11\% for one and 19\% for two undetected steps. However, further undetected steps provide no additional advantage to our approach, with the relative cost decreasing slightly to 18\% and 17\% for 3 and 4 undetected steps, respectively. The reason for the initial jump is clear, as the attacker holds more ground in the network and has more possibilities to pursue, and the PLE strategy loses its edge when there are choices to be made. The reduction to the benefit at higher number of undetected steps is due to an increase in average overall cost faced by both approaches, so the difference is smaller relative to this level.


Sensitivity of the results to different assumptions about the cost structure was also tested. These were again run for the 20-node graphs, and for delayed detection with 2 undetected steps, which provided the highest cost impact in the delay sensitivity tests. 
The tests consisted of varying two different settings relating to costs: the utility obtained from services per time unit, which affects the indirect costs arising from node unavailability; the ratio of the direct cost of recovery to the direct cost of patching, which can affect the cost-effectiveness of patching relative to recovery actions. 

\begin{figure}[h!]%
\centering
\includegraphics[width=\columnwidth]{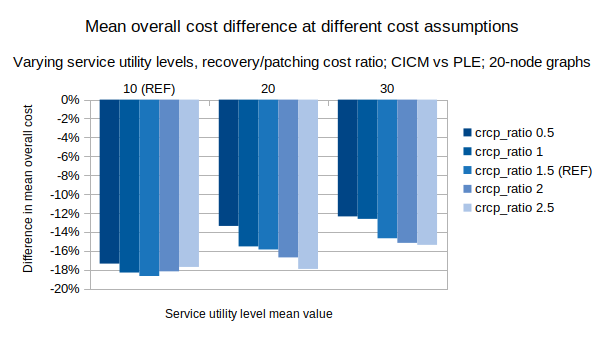}
\caption{Sensitivity to different cost parameter levels}
\label{fig:Cost_parameter_sensitivity_20-node}
\end{figure}

Fig. \ref{fig:Cost_parameter_sensitivity_20-node} shows the results for the cost parameter sensitivity.
We can see two broad patterns: First, the magnitude of the cost difference between the approaches is reduced as the average service utility level increases. Second, in most cases a higher cost of recovery (higher $c_R/c_P$ ratio) leads to an increase in the cost savings provided by CICM compared to PLE, other things equal.

The first observation suggests that the higher the potential loss from unavailability, the less room there is to find benefit from actions that proactively contains the attack as opposed to treating the latest compromise. Therefore, the cost-efficient approach becomes more similar to PLE, which not only treats the latest event but also patches only nodes that are already down, avoiding additional unavailability costs. However, the difference between the approaches is still sizeable, with the smallest difference in Fig. \ref{fig:Cost_parameter_sensitivity_20-node} suggesting a 12\% average saving using CICM relative to PLE. 

The second observation suggests that the higher the cost of recovery is relative to patching, the more room there is for CICM to find cost savings by deviating from treating the latest compromise, as the cost of taking proactive patching steps is reduced. The average cost levels of both approaches increases with the $c_R/c_P$ ratio, but the rise in the costs incurred is smaller when using CICM than PLE, leading to an increasing relative benefit. 

\section{Conclusion and Future Work} \label{sec:conclusion}
We proposed a framework for automated countermeasure selection based on cost impact analysis of the organisation's service loss and costs over a period of time considering both attacker and defensive actions, aiming for a cost-effective approach to maintaining service functionality.
The method was demonstrated via examples in a sample network, and an evaluation of its countermeasure selection performance was conducted using simulations. 
The results suggest that our method outperforms an alternative countermeasure selection approach based on the attack impact assessment method by \cite{Albanese2017,Albanese2011}, both in terms of average service performance and overall costs over a given time window. Comparisons against a straightforward patching approach showed that, while average service performance was a close match, our method found more cost-efficient ways to achieve the goal. 

Future work will involve adding more detail into the impact modelling, moving away from the simple DG as used here in favour of a more flexible dependency model. On the countermeasure selection side, we shall investigate extending the method to select a combination of countermeasures and recovery actions in the same decision. Additionally, we intend to look into the possibility for further improvements from POMDP-based methods to attack modelling.
We also intend to extend the method to consider more detailed disabling schemes based on network connectivity. The current model could estimate the impact of such schemes, but we do not currently have this additional topological information included in the attack or dependency graphs. Given this, we are yet to devise a method to evaluate the system's performance under such disabling schemes.

\bibliographystyle{ACM-Reference-Format}
\bibliography{references}


\end{document}